\documentclass[aps,prb,twocolumn,superscriptaddress,floatfix]{revtex4-2}

\usepackage{placeins}
\usepackage{graphicx}
\usepackage{amsmath}
\usepackage{amssymb}
\usepackage{dcolumn}
\usepackage[colorlinks=true,linkcolor=blue,citecolor=green,urlcolor=blue]{hyperref}

\begin{document}

\title{Magneto-transport and electronic structure studies of ternary antimonides, La\textit{T}Sb$_2$ (\textit{T} = Cu, Ag)}

\author{Himanshu Pant}
\affiliation{School of Physical Sciences, Indian Institute of Technology Mandi, Kamand, Himachal Pradesh- 175005, India}

\author{Neeraj Bhatt}
\affiliation{Department of Physics, Indian Institute of Science Education and Research, Bhopal, Madhya Pradesh 462066, India}

\author{Deepali Sharma}
\affiliation{Department of Physics, Indian Institute of Science Education and Research, Bhopal, Madhya Pradesh 462066, India}

\author{Asif Ali}
\affiliation{Department of Physics, Indian Institute of Science Education and Research, Bhopal, Madhya Pradesh 462066, India}

\author{A. Thamizhavel}
\affiliation{Department of Condensed Matter Physics and Materials Science, Tata Institute of Fundamental Research, Homi Bhabha Road, Colaba, Mumbai- 400005}

\author{Saurabh Singh}
\altaffiliation[Present adderess: ] {Quantum Dynamics Unit, Okinawa Institute of Science and Technology (OIST) Graduate University, Onna, Okinawa 904-0495, Japan}
\affiliation{Toyota Technological Institute, Nagoya, Aichi 468-8511, Japan}
\affiliation{Japan Science and Technology Agency, Kawaguchi, Saitama 332-0012, Japan}
\

\author{Kentaro Kuga}
\affiliation{Toyota Technological Institute, Nagoya, Aichi 468-8511, Japan}

\author{Tsunehiro Takeuchi}
\affiliation{Toyota Technological Institute, Nagoya, Aichi 468-8511, Japan}
\affiliation{Japan Science and Technology Agency, Kawaguchi, Saitama 332-0012, Japan}

\author{Ravi Shankar Singh}
\affiliation{Department of Physics, Indian Institute of Science Education and Research, Bhopal, Madhya Pradesh 462066, India}

\author{R. Bindu}
\email{bindu@iitmandi.ac.in}
\affiliation{School of Physical Sciences, Indian Institute of Technology Mandi, Kamand, Himachal Pradesh- 175005, India}

\date{\today}

\begin{abstract}

We investigate Sb square net-based Dirac semimetallic ternary antimonides La\textit{T}Sb$_2$ (\textit{T} = Cu, Ag) using magnetotransport, photoelectron spectroscopic studies, complemented by density functional theory (DFT) based ab initio calculations. Our results show that both the compounds show positive temperature coefficient of resistivity, linear magnetoresistance (MR) and deviation from Fermi liquid behaviour at low temperatures (2-30 K). But only \textit{T} = Ag compound exhibit signature of charge density wave (CDW) phenomena and the results of its temperature dependent MR data suggest that single-band semiclassical transport is not sufficient to understand the transport behaviour. $\text{DFT}+U$ studies show that in both samples, an on-site Coulomb interaction ($\textit{U}$) for Cu 3\textit{d} and Ag 4\textit{d} electrons, in the presence of spin-orbit coupling (SOC) is required to reproduce experimental valence band spectra. Sb 3\textit{d}, 4\textit{d} core-level spectra combined with $\text{DFT}+U$ calculations and $\text{La } 3\textit{d}$ spectra confirm the presence and effects of inequivalent Sb sites for both the samples. In contrast to the $\textit{T} = \text{Cu}$ compound, which follows expected temperature-dependent core-level sharpening, the $T = \text{Ag}$ compound exhibits an anomalous low-temperature broadening in the CDW phase. The analysis of Sb 3\textit{d} spectra suggests Sb square nets are linked to CDW phenomena in \textit{T} = Ag compound. Temperature-dependent ultraviolet photoemission spectroscopic studies on \textit{T} = Ag show a small decrement in spectral DOS at the $\epsilon_F$ for the temperature below the CDW transition temperature. We believe that our core-level studies on these compounds will be valuable to the scientific community by providing a comparative understanding of the electronic structures of CDW and non-CDW systems, clarifying the role of local electronic environments in CDW formation, and establishing a spectroscopic benchmark for future studies of related quantum materials.

\end{abstract}

%pacs{81.30.Kf,68.35.Gy,61.10.Nz,74.25.Jb}

\maketitle

\section{Introduction}

Various emergent properties have been observed in systems possessing a two-dimensional square lattice in their crystal structure. High T$_c$ superconducting cuprates\cite{sachdev2000quantum} and iron-pnictides\cite{fernandes2022iron} superconductors are important examples of this. In recent years, numerous intermetallic compounds have been established to be topological semimetals that have a square net in their crystal structure, showing various emergent properties. Rare earth-based ternary antimonides, RTSb$_2$, are systems that possess nearly all intriguing emergent features, making them an ideal platform to get some insights into the underlying physics. 

\begin{figure}

\includegraphics [width = 0.45\textwidth]{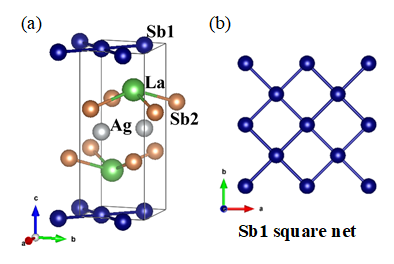}
\caption{(a) Crystal structure of LaTSb$_2$ (\textit{T} = Cu, Ag); (b) Sb1 square net.}
\label{Structure}

\end{figure}

In particular, Lanthanum ternary antimonides LaTSb$_2$ (\textit{T} = Cu, Ag) have attracted intense research interests due to their peculiar physical 
properties such as Dirac-cone-like band structure (Dirac-Fermion systems), charge density wave (CDW) order, linear 
magnetoresistance (MR) and superconductivity\citep{song2003XRD,singha2020,LaCuSb2_akiba2023,LaAgSb2_akiba2022}. LaTSb$_2$ (\textit{T} = Cu, Ag) samples crystallize in a simple tetragonal ZrCuSi$_2$-type structure (\textit{P4/mmm}m space group), in which La and T atoms are sandwiched in between Sb square nets\cite{song2003XRD}, shown in Fig.\ref{Structure}. There is one Wyckoff position for La(2c) and Cu(2b) atoms and two inequivalent Sb positions in the lattice structure, the first at 2a (Sb1) and the other at 2c (Sb2) Wyckoff positions.

According to literature reports, while LaCuSb$_2$ is a metal, LaAgSb$_2$ shows two different CDW transitions: (a) incommensurate CDW transitions around 211 K ($T_{C1}$) along \textit{a} direction and (b) an incommensurate CDW transition around 183 K ($T_{C2}$) which becomes commensurate around 164 K ($T_{C2-com}$) along \textit{c} direction\cite{song2003XRD}. These CDW phase transition temperatures can be manipulated using external pressure. Zhang et al.\cite{Zhang2021_Pc} showed destabilisation of CDW phase with applied pressure and a structural transition at $\sim$ 22 GPa while Singha et al.\citep{singha2020} have also reported multiple structural transitions with applied pressure using synchrotron and Raman studies. Using pressure-dependent resistivity measurements, Akiba et al.\citep{LAS_Akiba_2021} showed that the CDW transitions, $T_{C1}$ and $T_{C2}$ get totally suppressed at the critical pressures of P$_{C1}$ $\sim$ 3.2 GPa and P$_{C2}$ $\sim$ 1.7 GPa, respectively.

%Recently, in both samples, superconductivity has been confirmed, and according to combined experimental and DFT calculation results\cite{LaCuSb2_akiba2023, LaAgSb2_akiba2022}, authors have concluded that whereas the experimental superconducting transition temperature in LaCuSb$_2$ ($\sim$1 K) can be explained theoretically using the conventional framework of the phonon-mediated pairing mechanism, the superconducting transition temperature in LaAgSb$_2$ (0.29 K) can't be explained using this mechanism.

Recently, superconductivity has been confirmed in both samples. According to combined experimental results and DFT calculations\cite{LaCuSb2_akiba2023, LaAgSb2_akiba2022}, the authors concluded that the experimental superconducting transition temperature in LaCuSb$_2$ ($\sim$ 1 K) can be theoretically explained using the conventional framework of the phonon-mediated pairing mechanism. However, the superconducting transition temperature in LaAgSb$_2$ ($\sim$ 0.29 K) cannot be explained by this mechanism. This suggests that at ambient pressure, charge density waves in LaAgSb$_2$ significantly affect the physical properties at low temperatures. The superconducting transition temperature for LaAgSb$_2$ exhibits a maxima at the critical pressure P$_{C2}$\cite{LaAgSb2_akiba2022}. This transition is accompanied by a non-monotonic fluctuation of exponent n in $\rho$ = $\rho_{0}$ + AT$^{n}$, indicating a deviation from Fermi liquid behaviour near the quantum critical region\cite{kusmartseva2009pressure,gruner2017charge, LaAgSb2_akiba2022}. The value for exponent n $\sim$ 2.8 is reported at ambient pressure  for \textit{T} = Ag compound\cite{LaAgSb2_akiba2022}. In a recent publication, Alam et al. \cite{alam2023_PHE} have reported a very large planar Hall effect (PHE) in CDW phase below $T_{C2}$.

Within the $\text{LaTSb}_2$ ($T = \text{Cu, Ag, Au}$) series, a pronounced competitive interplay between superconductivity (SC) and CDW order is observed. The superconducting transition temperature reaches its maximum in the $T = \text{Cu}$ compound ($\sim 1\text{ K}$, where CDW ordering is entirely absent \cite{LaCuSb2_akiba2023}. Conversely, the $T = \text{Ag}$ compound exhibits the lowest SC transition temperature ($\sim 0.3\text{ K}$) alongside the highest CDW transition temperature ($T_{\text{CDW}} = 207\text{ K}$)\cite{LaAgSb2_akiba2022}, while the $T = \text{Au}$ compound sits intermediately with moderate SC transition temperature ($\sim 0.6\text{ K}$) and $T_{\text{CDW}}$ values ($\sim 88\text{ K}$)\cite{kuo2019characterization}. Interestingly, the unit cell volume displays an inverse correlation with SC transition temperature, shrinking systematically from $209.84\text{ \AA}^3$ ($T = \text{Ag}$)\cite{LAS_Akiba_2021} to $205.80\text{ \AA}^3$ ($T = \text{Au}$)\cite{lingannan2021pressure} and $198.1\text{ \AA}^3$ ($T = \text{Cu}$)\cite{LCS_Sologub_1994}, which suggests that lattice compression suppresses the CDW phase in favour of superconductivity. The results on pressure-dependent studies on LaAgSb$_2$ also suggest the fact that increasing pressure increases the superconducting transition temperature to $\sim$1 K at critical pressure.

Core-level photoemission spectroscopy provides vital information about element-specific electronic structure. By understanding the well-screened and poorly screened features and shake-up satellites, one can disentangle charge-transfer dynamics, many-body interactions, and local bonding environments. In the case of CDW systems, especially when there is a commensurate charge density wave, core level spectroscopy serves as an important technique to capture its signatures through the splitting of the respective core levels. The best example of it is model CDW systems 1T-TaS$_2$\cite{ishizaka2011femtosecond, hughes1976charge} and 1T-TaSe$_2$\cite{horiba2002charge, hughes1976charge}, where CDW charge redistribution can be clearly seen in the Ta 4f core level spectra as the distinct core level splitting.

Complementing this real-space insight, angle-resolved photoemission spectroscopy (ARPES) has also emerged as an indispensable technique for mapping reciprocal space. Many metallic CDW systems, such as $\text{RTe}_3$\cite{brouet2008angle}, $2\text{H-TaSe}_2$ \cite{borisenko2008pseudogap}, $1\text{T-TaS}_2$ and $1\text{T-TaSe}_2$ \cite{bovet2004pseudogapped}, exhibit partial gapping or a pseudogapped Fermi surface as a characteristic fingerprint of the CDW phase.

In the available literature, photoemission experiments have been used to uncover rich electronic properties in LaTSb$_2$ (\textit{T} = Cu, Ag). In the CDW-ordered compound $\text{LaAgSb}_2$, Shi et al. \cite{ARPES_shi2016} reported a Dirac cone-like band structure in the vicinity of the Fermi level ($\epsilon_F$) alongside nested Fermi surface pockets. Corresponding core-level spectra further revealed a distinct splitting ($\sim$0.2 eV) in the Sb 4\textit{d} core level, arising from the two inequivalent Sb sites. Subsequent high-resolution ARPES by Rosmus et al. \cite{ARPES_rosmus_LAS} highlighted the presence of highly degenerate nodal lines located at the boundary of the Brillouin zone and emphasised the role of surface termination in observing these linear dispersions. Interestingly, when turning to the structurally isomorphic but non-CDW compound $\text{LaCuSb}_2$, ARPES investigations \cite{ARPES_rosmus_LCS, Chamorro_LCS_2019} also revealed the presence of Dirac-like bands and nodal lines. Crucially, Rosmus et al. \cite{ARPES_rosmus_LCS} pointed out that the experimentally observed nesting of the exterior Fermi surface branches actually appears more pronounced in $\text{LaCuSb}_2$ than in $\text{LaAgSb}_2$. This observation stands contrary to conventional expectations given that CDW transitions manifest exclusively in the $\text{LaAgSb}_2$, posing an intriguing puzzle regarding the driving mechanism of the CDW. In LaAgSb$_2$, Chainani et al.\cite{chainani2025_LAS}, in their temperature-dependent hard x-ray photoemission spectroscopy (HAXPES) study have reported a well-separated satellite below the CDW transition temperature $T_{C1}$ in the Sb 3\textit{d} core level. This satellite shows a progressively increasing intensity below the $T_{C2}$. To the best of our knowledge, there are no core-level studies have been reported on the \textit{T} = Cu system in the literature.

Keeping all these in mind, we have carried out magnetotransport, core level, and valence band X-ray photoemission (XPS) and ultraviolet photoemission (UPS) spectroscopic studies combined with the electronic structure calculations using $\text{DFT}+U$ methodology. Transport measurements show positive temperature coefficient of resistivity (TCR), direction-dependent linear Magnetoresistance (MR) (more pronounced in \textit{T} = Ag) and deviation from Fermi liquid behavior (2-30 K) for both the compounds. However, signatures of charge density wave (CDW) ordering is observed only for \textit{T} = Ag compound. Temperature-dependent MR measurements for \textit{T} = Ag, show change of slope from linear to quadratic behaviour as temperature increases, i.e., as CDW critical temperature is reached. Deviation from Kohler's scaling indicates that single-band semiclassical transport is not sufficient for explaining the transport behaviour in \textit{T} = Ag system. $\text{DFT}+U$ calculations are required for better representation of the valence band spectra for both the compounds. High-quality core-level data confirms the existence of inequivalent Sb sites in the Sb 3\textit{d} and 4\textit{d} spectra. The influence of these inequivalent Sb sites is evident in the La 3\textit{d} spectra as well. The existence of inequivalent Sb sites was also verified using DFT+U calculations for Sb 4\textit{d} spectra for both the systems. Sb 3\textit{d} core-level studies also hint towards the Sb square net being linked with the CDW phenomena for \textit{T} = Ag system. Unlike the $T = \text{Cu}$ system, which demonstrates anticipated temperature-dependent core-level sharpening at low temperatures due to reduced phonon scattering, the $T = \text{Ag}$ compound displays an unusual low-temperature broadening in the CDW phase. Detailed UPS studies (He-I spectra) on the $T = \text{Ag}$ compound show a small decrement in spectral DOS at the $\epsilon_F$ below the CDW transition temperature.

\section{Experimental}

High-quality single crystals of LaTSb$_2$ (T= Cu/Ag) were grown using Sb as a self-flux by the high-temperature solution growth
method\cite{Thamizhavel_2003}. 

%Highly pure constituent elements La, Cu/Ag and Sb were taken in the molar ratio of 1:2:21 in a high-quality recrystallised alumina crucible; here, excess Cu/Ag was used to avoid the LaSb2 phase formation. The crucible containing the elements was sealed in an evacuated quartz ampoule with a partial pressure of Ar gas. The crucible was then placed in a box-type resistive heating furnace. The heating was done slowly at the rate of 50 $^o$C/h up to 1050 $^o$C and held at this temperature for about 24 hours for proper homogenisation. Then the furnace was cooled down at the rate of 2 $^o$C/hr down to 700 $^o$C, at which point the excess Sb flux was centrifuged. Well defined shiny single crystals with platelet-like morphology were obtained. The flat plane of the crystal corresponds to the (001) plane. 

The temperature-dependent photoemission spectra were collected on a Scienta R4000 hemispherical analyser, using monochromatic Al K$_\alpha$ (1486.6 eV), He-I (21.2 eV) and He II$^*$ (48.37 eV) radiation sources at 30 K and 300 K. The base pressure in the main analysis chamber is of the order of 10$^{-11}$ mbar during the experiment. Total instrumental resolutions were set to $\sim$ 300 meV $\sim$ 5 meV and $\sim$ 15 meV for Al K$\alpha$, He I and He II$^*$ radiations, respectively. The sample surface was cleaned in situ via cleaving. The binding energy was calibrated by measuring the $\epsilon_F$ of an Ag pellet cleaned in situ by argon ion sputtering.

The temperature-dependent resistivity and magnetoresistance measurements using the standard four-probe method were carried out on square single-crystal pieces using the Physical Properties Measurements System from Quantum Design, USA.

%He-II (40.8 eV) 

\section{Computational Details}

Density of states (DOS) calculations were performed using a full potential linearised augmented plane wave basis with local orbitals (LAPW + lo) as implemented in Elk code\cite{dewhurst2011_Elkcode}. For the exchange and correlation part, we have employed generalised gradient (GGA)- PBEsol approximation\cite{perdew2008_PBEsol}. On site Coulomb interaction term (\textit{U}) was used to take into account the strong electron-electron correlation in Ag 4\textit{d} and Cu 3\textit{d} orbitals. The experimental lattice parameters and the atomic positions were fully relaxed starting from Akiba et al.\citep{LAS_Akiba_2021} and Sologub et. al.\cite{LCS_Sologub_1994}. The muffin-tin sphere radii of  2.8, 2.6, and 2.6 bohr were used for La, Ag, and Sb, respectively, in the case of LaAgSb$_2$ and 2.8, 2.35, and 2.55 bohr were used for La, Cu, and Sb, respectively, in the case of LaCuSb$_2$. For the termination of the self-consistent cycle, the difference in  total energy was set to be less than 10$^{-4}$ Hartree/unit cell in the calculations. The structure was considered to be relaxed when the total force was less than 0.5$\times$10 $^{-3}$ Hartree/unit cell. These calculations were performed using a 13$\times$13$\times$6 k-point mesh for all the calculations.

\section{Results and discussion}

\subsection{Magnetotransport Studies}

\begin{figure}

\includegraphics [width = 0.5\textwidth]{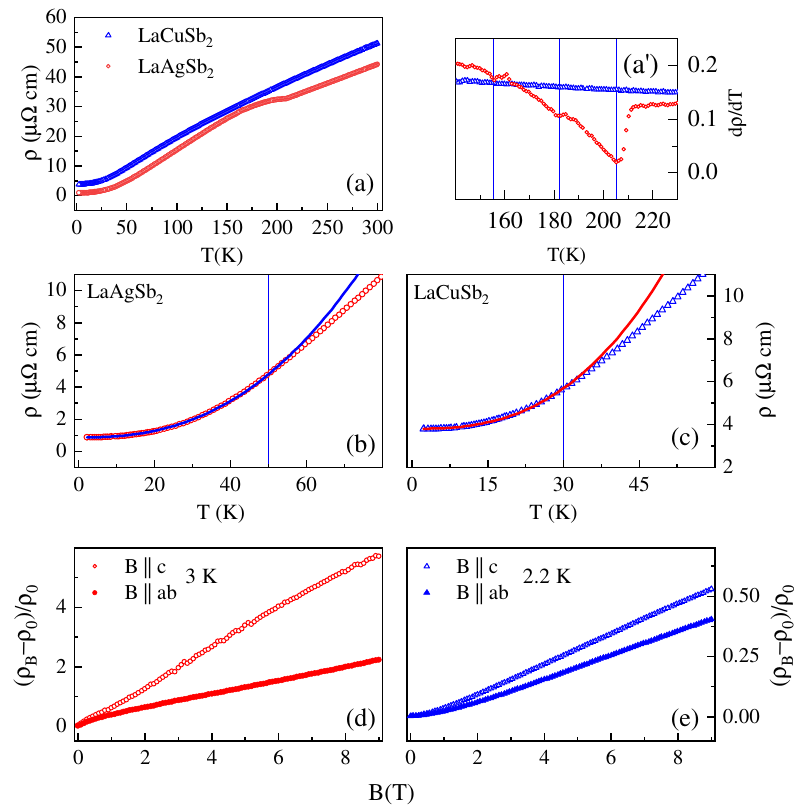}
\caption{(a) Temperature dependent resistivity for LaAgSb$_2$ and  LaCuSb$_2$; (a') the first derivative of resistivity; Low temperature resistivity fitting for (b) LaAgSb$_2$; (c) LaCuSb$_2$; transverse and linear magnetoresistence for (d) LaAgSb$_2$ at 3 K; (e) LaCuSb$_2$ at 2.2 K.}
\label{resistivity}

\end{figure}

Fig.\ref{resistivity}(a) shows the temperature-dependent resistivity behaviour (I $\parallel$ \textit{ab}, i.e., $\perp$ to the growth axis) for LaAgSb$_2$ and LaCuSb$_2$ single crystals collected in the temperature range 300 to 2.2 K. Both the samples show an overall decrease with the decrease in temperature, but a hump-like feature near 200 K is observed for the case of LaAgSb$_2$. To look more closely for the changes in the resistivity data, we have taken the first derivative of temperature-dependent resistivity data, shown in Fig. \ref{resistivity}(a'). We observed one sharp downturn around $\sim$ 205 K and two small features around $\sim$ 182 K and $\sim$ 155 K for LaAgSb$_2$ and no change for LaCuSb$_2$ in the 140 - 230 K temperature range. As previously reported in literature\cite{song2003XRD,singha2020}, the hump-like feature in LaAgSb$_2$ is related to CDW transition. LaAgSb$_2$ shows two CDW transitions, $T_{C1}$ $\sim$ 211 K and $T_{C2}$ $\sim$ 183 K. The first incommensurate transition occurring at $\sim$ 211 K, develops along \textit{a} direction involving $\sim$ 40 unit cells and the second incommensurate transition occurring at $\sim$ 180 K, develops along \textit{c} direction, which turns commensurate at a lower temperature, $\sim$ 160 K, involving 6 unit cells. We were able to observe the signature of these transitions in $d\rho/dT$ vs T data, occurring around $\sim$ 205 K, $\sim$ 182 K and $\sim$ 155 K, respectively, Fig. \ref{resistivity}(a'). The metallic behaviour observed in case of LaCuSb$_2$ without any transitions is also in line with the previously reported literature\cite{LaCuSb2_akiba2023}.

To understand the behaviour of the low-temperature resistivity data of both the compounds, fitting, Figs. \ref{resistivity}(b) and (c), was carried out using the functional form as given below 

\begin{equation*}
 \rho = \rho_{0} + A T^{n} 
 \label{FL}
\end{equation*}

Our results show that the value of n obtained is 2.65(2) and 2.46(2) in the case of LaCuSb$_2$ and LaAgSb$_2$, respectively. As the value of n is  not equal to 2, this suggests that both the compounds are not following the Fermi-liquid theory at low temperatures. The obtained value of n $>$ 2 is in line with the pressure-dependent studies carried out on LaAgSb$_2$, where the value for exponent n $\sim$ 2.8 is reported at ambient pressure for the \textit{T} = Ag compound\cite{LaAgSb2_akiba2022}. The value of this exponent shows anomalous behaviour at the CDW quantum critical point, as discussed in the introduction.  It appears that the origin of such a functional form requires more theoretical investigation.

%\textbf{However, in several 3D magnetic quantum critical metals (Ce(Ni$_{0.935}$Pd$_{0.065}$)$_2$Ge$_2$\cite{wang2015quantum}, URhAl\cite{shimizu2015unusual}, U$_3$Ni$_5$Al$_{19}$\cite{bauer2005high}) with exponent (n$<$2), fractional exponents such as n=1.5 (T$^{3/2}$), near AFM quantum critical points (QCPs), and n=1.67 (T$^{5/3}$), near the ferromagnetic QCPs, are observed, which are broadly consistent with Hertz-Millis-Moriya (HMM) spin-fluctuation theory. Whereas for some other systems having exponent n$<$2, such as systems with linear in T (n=1) behaviour (Cuprates\cite{ayres2021incoherent}, BaFe$_2$(As$_{(1-x)}$P$_x$)$_2$\cite{shibauchi2014quantum}, YBRh$_2$Si$_2$\cite{gegenwart2002magnetic}, CsV$_3$Sb$_5$\cite{feng2023commensurate}), a universal microscopic theory predicting exact power-law exponents near charge density wave (CDW) instabilities and superconducting domes still remains an active open problem.} 

%At higher temperatures ($>$250 K), above the Debye temperature for the materials, both the samples show linear behaviour, which indicates that above this temperature phonons are dominating the transport properties in these materials. This is commonly observed behaviour in the case of metals and metallic alloys[].

These compounds host robust Dirac-like bands due to their non-symmorphic structure, which can affect electronic transport in the presence of a magnetic field; Hence, to examine the effect of the magnetic field on transport behaviour, we have performed low-temperature magnetotransport experiments on both compounds. Figs. \ref{resistivity}(d) and (e) show the magnetoresistance (MR) for LaAgSb$_2$ in transverse (B $\parallel$ \textit{c}) and longitudinal (B $\parallel$ \textit{ab}) configurations. For both the configurations, the observed MR is non-saturating until an applied magnetic field of 9 T at $\sim$ 3 K. The value of transverse MR is 2.6 and 1.3 times greater than longitudinal MR for LaAgSb$_2$ and LaCuSb$_2$, respectively, at a 9 T applied magnetic field. Although both the compounds show non-saturating linear MR at low temperatures and high fields, the absolute value for transverse MR is almost ten times higher for LaAgSb$_2$ in comparison to LaCuSb$_2$. This shows that LaAgSb$_2$ exhibits a much stronger directional dependence of charge transport than LaCuSb$_2$.

%specifically, systems hosting Dirac and Weyl fermions can exhibit extremely high and non-saturating linear magnetoresistance at high fields. 

\begin{figure*}

\includegraphics [width = 0.80\textwidth]{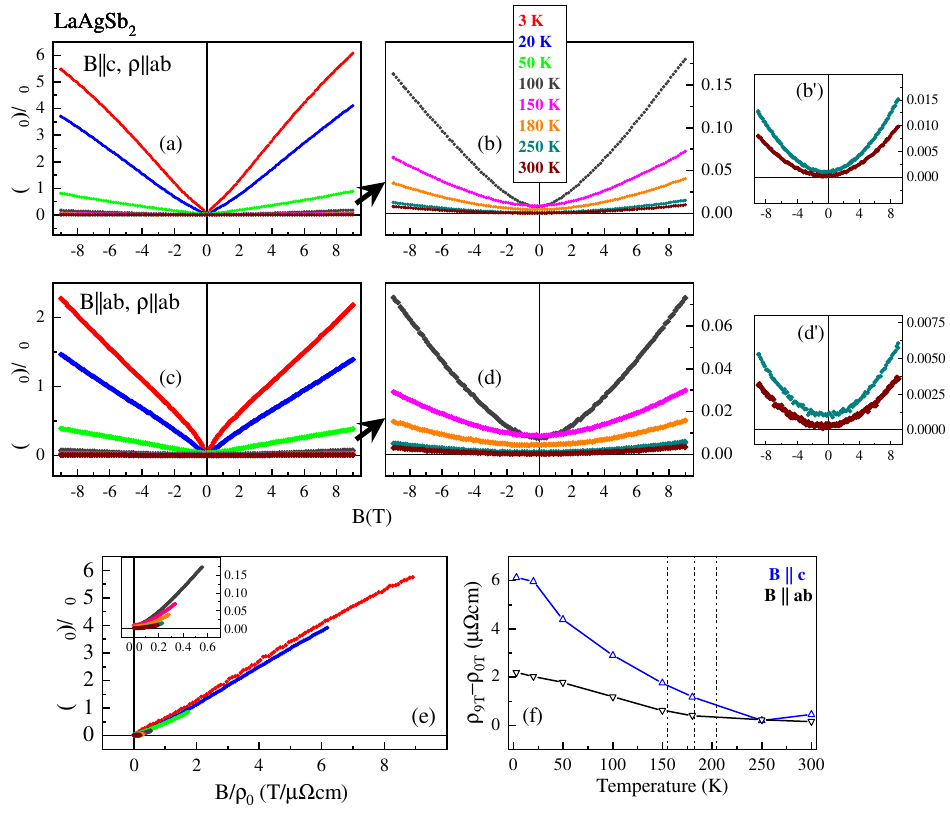}
\caption{Temperature-dependent magnetoresistance for LaAgSb$_2$ in (a),(b),(b') transverse configuration; (c),(d),(d') longitudinal configuration; (e) Kohler's plots for transverse MR; the inset shows the Kohler's plot for higher temperatures; (f) resistivity data with and without a field; for the with-field data (black and blue), lines are a guide to the eye; (f') temperature-dependent magnetoresistance at 9T applied field; black and blue lines are a guide to the eye. In transverse configuration, the magnetic field is applied in the plane perpendicular (c-plane) to current (ab-plane), and longitudinal configuration is when the magnetic field is in the plane parallel to the current.}
\label{MR_T}

\end{figure*}

To further investigate the nature of transport in LaAgSb$_2$, more detailed MR studies have been carried out. Figs. \ref{MR_T}(a-d) show the transverse and longitudinal MR data for different temperatures; these results are in line with the previous report by Wang et. al.\cite{Wang2012}. We observe an asymmetry between the positive and negative cycles of the MR curves, Figs. \ref{MR_T}(a-d). This mismatch in the data is possibly arising due to the large Hall effect observed in the sample\citep{Wang2012}. To nullify its effect, we have taken the average for MR curves in both cycles and used that data for further analysis.

 In a single band semi classical transport, the MR scales as a function of B$\tau$, where B is the applied magnetic field and $\tau$ is the relaxation time. With the assumption that there is only one single scattering time, 1/$\tau$(T) $\propto$ $\rho_0$(T), where $\rho_0$(T) is the zero field resistivity\cite{Wang2012}, then all the MR curves at different temperature should overlap if plotted against B/$\rho_0$(T). This is known as Kohler's scaling rule. To verify Kohler's scaling rule, in Fig. \ref{MR_T}(e), we plotted MR vs B/$\rho_0$. Our results show that the transverse MR deviates from Kohler's scaling rule. This suggests that single-band semi classical transport is not sufficient to explain the transport behaviour in LaAgSb$_2$.

Our results show that at low temperatures MR shows almost linear behaviour, as shown in Figs. \ref{MR_T}(a) and (c). As we move towards higher temperatures, MR behaviour moves towards quadratic dependence, as shown in Figs. \ref{MR_T}(b) and (d). To draw some idea of the field-dependent behaviour of resistivity, we have plotted the available field-dependent resistivity data subtracted from the resistivity data in no applied field in both the configurations at a 9 T field, Fig. \ref{MR_T}(f). It is interesting to note that below the CDW transition temperatures (shown by dashed lines in the graph) we observe an anomalous increase in MR for both configurations.

Several mechanisms can account for linear magnetoresistance (MR) in condensed matter systems. While linear MR commonly arises as an average effect in polycrystalline samples, this can be ruled out here, as high-quality single crystals were studied. Alternatively, linear MR can be realized in the quantum limit for materials hosting linear Dirac or Weyl bands. Under a sufficiently high magnetic field, the electronic structure quantizes into Landau levels (LLs). When the magnetic field is large enough that the LL separation ($\Delta_{LL}$) between 0$^{th}$ and 1$^{st}$ LL exceeds both the Fermi energy ($E_F$) and the thermal fluctuation energy ($k_B T$), all carriers occupy the lowest ($0\text{th}$) LL. For conventional parabolic bands, this quantum limit is typically inaccessible at moderate magnetic fields ($< 9\text{ T}$). However, for materials with linearly dispersing Dirac bands, the unique LL spacing ($\Delta_{LL} \propto \sqrt{B}$) allows the system to reach the quantum limit at much lower, moderate fields, inducing a robust linear MR as described by Abrikosov's model\cite{abrikosov1998quantum,abrikosov2000quantum,Wang2012}.

Linear MR can also occur in some materials with charge or spin density waves\cite{feng2019linear}. When these density waves form, they alter the electronic structure of the material, breaking the Fermi surface into small pockets with highly curved, sharp corners. Under a magnetic field, electrons are forced into circular orbits. As the magnetic field increases, the electron speed (cyclotron frequency) increases proportionally. This allows a linearly increasing number of electrons to successfully navigate these sharp corners within their scattering lifetime so that the electrical transport becomes more difficult, resulting in a steady linear MR response even at very low magnetic fields\cite{feng2019linear}.

In the case of both the compounds, as the Dirac like linear band dispersion in the band structure is well established, the observation of linear MR at low temperature can be understood due to the quantum limit mechanism. This is especially relevant for $\text{LaCuSb}_2$, given the absence of a CDW in this system. Interestingly, $\text{LaAgSb}_2$ exhibits a significantly higher MR magnitude, Fig. \ref{resistivity}(e), along with an anomalous increase in MR below the CDW transition temperature, Fig. \ref{MR_T}(f'). This indicates that for the case of LaAgSb$_2$ the Dirac bands and the CDW-reconstructed Fermi surface work cooperatively to dictate the low-temperature linear MR behaviour \cite{Wang2012, feng2019linear}. Nevertheless, detailed magneto-transport studies coupled with precise modeling of the CDW state in $\text{LaAgSb}_2$ remain essential to fully map the reconstructed Fermi surface and clarify its exact impact on transport.

Now that we have characterised both samples using detailed transport studies, temperature-dependent x-ray and ultraviolet (UV) photoemission studies have been conducted to gain a better understanding of the electronic structure in these samples.

%In the case of LaAgSb$_2$, we believe the presence of Dirac bands and CDW are both working cooperatively in defining the linear MR behaviour at low temperature \cite{Wang2012, feng2019linear}; that is why we are observing more asymmetric and higher values of linear MR for LaAgSb$_2$ in comparison to LaCuSb$_2$, Figs. \ref{resistivity} (d), (e), and \ref{MR_T} (f), (f'). However, detailed magneto-transport studies along with proper modelling of CDW in LaAgSb$_2$ will be essential to study the Fermi surface in the CDW phase and its effect on transport.

%Figs. \ref{Ang_MR}(a)-(c), shows the angular response of MR for LaAgSb$_2$. Here 0$^{\circ}$ and 90$^{\circ}$ corresponds to magnetic field $\perp$ and $\parallel$ to the current (I $\parallel$ \textit{ab})in the sample in the 9 T applied field. This data is in line with the data recently reported by Akiba et. al.\cite{Akiba2022_MR}. It shows a dumb bell shape (two fold symmetry) in the MR distribution for the sample. The angular magnetoresistance (AMR) is primarily dependence on the symmetry of the cross-section of Fermi surface  perpendicular to the direction of current flow\cite{Akiba2022_MR}, the twofold symmetry is expected considering the tetragonal crystal structure. Our data confirms that the conclusion about the dumbbell-shaped AMR resulting from the tetragonal structure is valid, as the AMR consistently exhibits a dumbbell shape throughout the CDW transition up to 300 K. However, it is interesting to note that the low-temperature behaviour with the CDW phase is more quantitatively pronounced. 

\subsection{Valence band Spectra}

\begin{figure*}

\includegraphics [width = 0.80\textwidth]{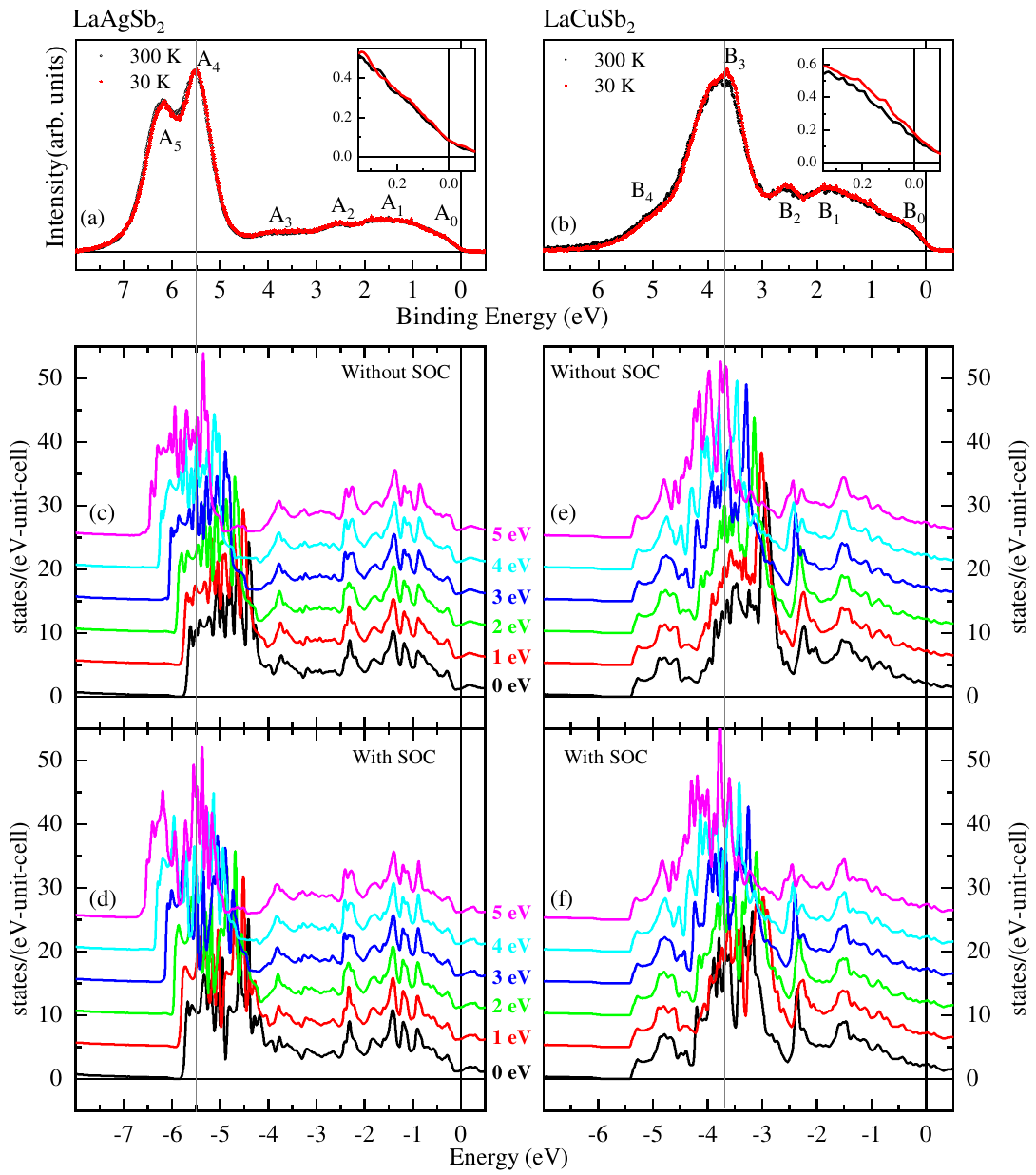}
\caption{Comparison of xps valence band spectra at 300 K and 30 K for LaTSb$_2$ (a) \textit{T} = Ag and (b) \textit{T} = Cu. Total density of states (TDOS) with varying $\textit{U}$ value for (c) \textit{T} = Ag without inclusion of spin orbit coupling (SOC); (d) \textit{T} = Ag with SOC; (e) \textit{T} = Cu without SOC and (f) \textit{T} = Cu with SOC. In panels (a) and (b) the spectra are normalised by dividing by the area under the curve following the background subtraction.}
\label{Valence}

\end{figure*}

\begin{table}

\setlength{\tabcolsep}{1.5pt}
  \renewcommand{\arraystretch}{1.2}
  
  \caption{Experimental and relaxed structural parameters for LaTSb$_2$. Wyckoff positions - La 2c, Ag 2b, Sb1 2a, Sb2 2c.}
  
  \begin{tabular}{lc|c|c|c}
  
   \hline
  &\multicolumn{2}{c}{\textbf{\textit{T} = Ag}} & \multicolumn{2}{c}{\textbf{\textit{T} = Cu}}\\
  \hline
  \hline
  &Experimental & Relaxed & Experimental & Relaxed\\     
  \hline
  
  a=b(\AA)&4.3940&4.3624&4.3690&4.3469\\
  c (\AA) &10.8680&11.0267&10.3760& 10.4817\\
  z$_{La}$  &0.2396&0.2416&0.2529&0.2531\\
  z$_{Sb2}$  &0.3304&0.3388&0.3524&0.3587\\
  
  \hline
  
   \multicolumn{5}{l}{Bond Distances}\\ 
   
   \hline
   
  Sb 2 - T (\AA) &\multicolumn{2}{c}{2.81} & \multicolumn{2}{c}{2.62}\\
  Sb 1 - Sb1 (\AA) &\multicolumn{2}{c}{3.084} & \multicolumn{2}{c}{3.073}\\
  La - Sb2 (\AA) &\multicolumn{2}{c}{3.266} & \multicolumn{2}{c}{3.267}\\
  La - Sb1 (\AA) &\multicolumn{2}{c}{3.443} & \multicolumn{2}{c}{3.429}\\

  \hline

  \end{tabular}
  \label{I}

\end{table}

Figs. \ref{Valence}(a) and (b) show xps valence band (VB) spectra at 300 K and 30 K for \textit{T} = Ag and \textit{T} = Cu, respectively. In the VB spectra of \textit{T} = Ag, broadly six features are visible, A$_0$ - A$_5$, respectively, shown in Fig. \ref{Valence} (a).  Similar to \textit{T} = Ag, in \textit{T} = Cu VB spectra, five features have been identified, B$_0$ - B$_4$, respectively, shown in Fig. \ref{Valence}(b).

There is no apparent change in VB features between 300 K and 30 K for both the samples. The insets of Figs.\ref{Valence}(a) and (b) show the temperature-dependent behaviour of intensity close to $\epsilon_F$. Negligible change is observed with change in temperature in the VB spectra near the $\epsilon_F$ for \textit{T} = Ag, shown in the inset of Fig. \ref{Valence}(a), while a small increment in intensity in the low-temperature spectra for \textit{T} = Cu can be seen near $\epsilon_F$, shown in the inset of Fig. \ref{Valence}(b).

To investigate the contribution of different electronic states in the valence band, first-principles density functional theory (DFT) calculations were performed for $\text{LaTSb}_2$ ($T = \text{Ag, Cu}$). In the case of $\text{LaAgSb}_2$, standard DFT provides an accurate representation of features $\text{A}_0$ to $\text{A}_3$, but a mismatch occurs for the high-intensity features $\text{A}_4$ and $\text{A}_5$ arising from Ag 4\textit{d} states\citep{ARPES_shi2016}, as shown in Figs. \ref{Valence} (c) and (d). To resolve this, $\text{DFT}+U$ calculations were conducted by varying \textit{U} from $1$ to $5\text{ eV}$ on the Ag 4\textit{d} orbitals without and with the inclusion of spin-orbit coupling (SOC). Increasing $\textit{U}$ smoothly shifts the TDOS corresponding to $\text{A}_4$ and $\text{A}_5$ to higher binding energies with minimal changes to the states near the $\epsilon_F$. We observe that the calculated spectra with the inclusion of SOC matches better with the experimental spectra. The experimental valence band spectra are satisfactorily reproduced using the calculation of $U = 5\text{ eV}$ including SOC.

Similarly, for LaCuSb$_2$, the standard $\text{DFT}+U$ fails to reproduce the $B_3$ feature correctly, which has a major contribution from localised Cu 3\textit{d}, as shown in Figs. \ref{Valence} (e) and (f). To account for electron-electron correlations, {$\text{DFT}+U$} calculations were performed with $\textit{U}$ values ranging from 1 to 5 eV. Without SOC, the $B_3$ feature gradually shifts to higher binding energy as $\textit{U}$ increases. However, with the inclusion of SOC, the feature initially shifts to lower binding energy at $\textit{U}$ = 1 eV before shifting toward higher binding energy. Similar to the Ag system, the $\textit{U}$ = 5 eV calculation with SOC provides the best agreement with the experimental valence band spectra. The necessity of a large $\textit{U}$ value to reproduce the experimental valence band spectra demonstrates the importance of strong electron-electron correlations in defining the electronic structure of this 3\textit{d} system.

\begin{figure}

\includegraphics [width = 0.48\textwidth]{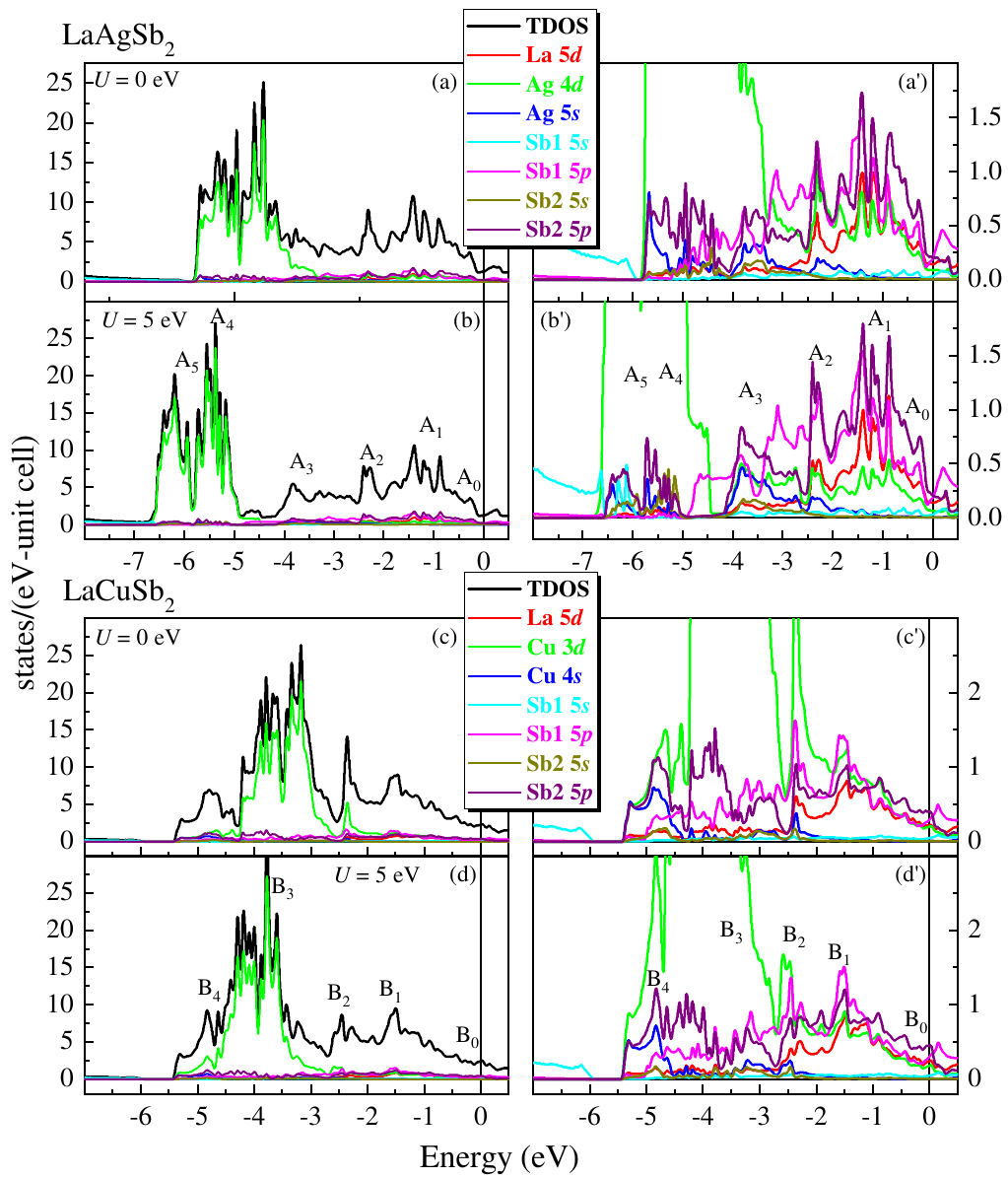}
\caption{Total and partial DOS for LaTSb$_2$ (a) \textit{T} = Ag with $\textit{U}$ = 0 eV and with SOC, (a') expanded view of the same; (b) \textit{T} = Ag  with $\textit{U}$= 5 eV and with SOC, (b') expanded view of the same; (c) \textit{T} = Cu with $\textit{U}$ = 0 eV and with SOC, (c') expanded view of the same; (d) \textit{T} = Cu  with $\textit{U}$= 5 eV and with SOC, (d') expanded view of the same.}
\label{DFT}

\end{figure}

The partial density of states (PDOS) from $\text{DFT}+U$ calculations with $\textit{U}$ = 0 eV and 5 eV with spin-orbit coupling (SOC) for both compounds are shown in Fig. \ref{DFT}. Figs. \ref{DFT}(a) and (a') show the results of $\text{DFT}+U$ calculations without the inclusion of $\textit{U}$ for \textit{T} = Ag. It can be readily observed that features A$_0$ - A$_3$ have major contributions coming from Sb 5\textit{p}, La 5\textit{d}, and Ag 4\textit{d} states along with almost negligible contributions from Ag 5\textit{s} and Sb 5\textit{s} states. On the other hand, features A$_4$ and A$_5$ have major contributions from Ag 4\textit{d} states and weak contributions from Sb 5\textit{p} and Ag 5\textit{s} states. With the inclusion of U (=5 eV), the position of Ag 4\textit{d} states (A$_4$ and A$_5$) shifts to the experimental B.E. position, Fig.\ref{DFT}(b), but on a closer look, it is also evident that the shape of the TDOS is more in line with the experimental spectra. It is important to note that, with the inclusion of $\textit{U}$ = 5 eV, all the states contributing close to $\epsilon_F$ exhibit similar behaviour to that for U = 0 eV. However, unlike the situation for $\textit{U}$ = 0 eV, in the energy range of 4 to 5 eV below $\epsilon_F$, there is a minor contribution from Sb1 (square net) 5\textit{p} states, in addition to the major contribution from Ag 4\textit{d} states, for $\textit{U}$ = 5 eV.

To summarise, features A$_0$, A$_1$, and A$_2$ have major contributions in the following order: Sb2 5\textit{p} $>$ Sb1 5\textit{p} $>$ La 5\textit{d} $>$ Ag 4\textit{d}, while they have a minor contribution in the following order: Ag 5\textit{p} (not shown) $>$ La 6p (not shown) $>$ Ag 5\textit{s} $>$ (Sb1 5\textit{s}, Sb2 5\textit{s}) states, as shown in Fig. \ref{DFT}(b'). Feature A$_3$ has a lower contribution from La 5\textit{d} states but a higher contribution from Ag 5\textit{s} states than features A$_0$, A$_1$, and A$_2$. Features A$_4$ and A$_5$ has dominant contribution from Ag 4\textit{d} and weak contributions from Sb2 5\textit{p}, Ag 5\textit{s}, Sb1 5\textit{s} and Sb2 5\textit{s} and Sb1 5\textit{p}, respectively.

Now let's discuss the VB contributions in case of \textit{T} = Cu. Figs. \ref{DFT}(c) and (c') show the results of DFT calculations without inclusion of $\textit{U}$ for \textit{T} = Cu.  It can be readily observed that features B$_0$ and  B$_1$ have major contributions from Sb 5\textit{p}, La 5\textit{d} and Cu 3\textit{d} states, while negligible contribution from Sb 5\textit{s} states, and features B$_2$, B$_3$ and B$_4$ have dominant contributions coming from Cu 3\textit{d} states and weaker contributions from Sb 5\textit{p} La 5\textit{d} and Cu 5\textit{s} states. Moreover, the inclusion of $\textit{U}$ = 5 eV in the Cu 3\textit{d} states gives better representation of the experimental spectrum in terms of the intensity ratio of feature B$_1$ to B$_2$ and also the energy position of feature B$_3$.  

In summary, features B$_0$ and B$_1$ have major contributions from different states in the following order: Sb2 5\textit{p} $>$ Sb1 5\textit{p} $>$ La 5\textit{d} $>$ Cu 3\textit{d} states, while they have negligible contribution in the following order from La 6\textit{p} (not shown) $>$ Cu 4\textit{p} (not shown) $>$ Cu 4\textit{s} $>$ (5\textit{s} of Sb1 and Sb2) states, shown in Fig. \ref{DFT}(d'). Feature B$_2$ has comparatively higher contributions from Cu 3\textit{d} and Cu 5\textit{s} states in comparison to B$_0$ and B$_1$. The features B$_3$ and B$_4$ have major contributions in the following order: Cu 3\textit{d} $>$ Sb2 5\textit{p} $>$ Sb1 5\textit{p} $>$ Cu 4\textit{s} and negligible contributions from Sb 5\textit{s}, La 6\textit{p} (not shown) and Cu 4\textit{p} (not shown).

\begin{figure}

\includegraphics [width = 0.48\textwidth]{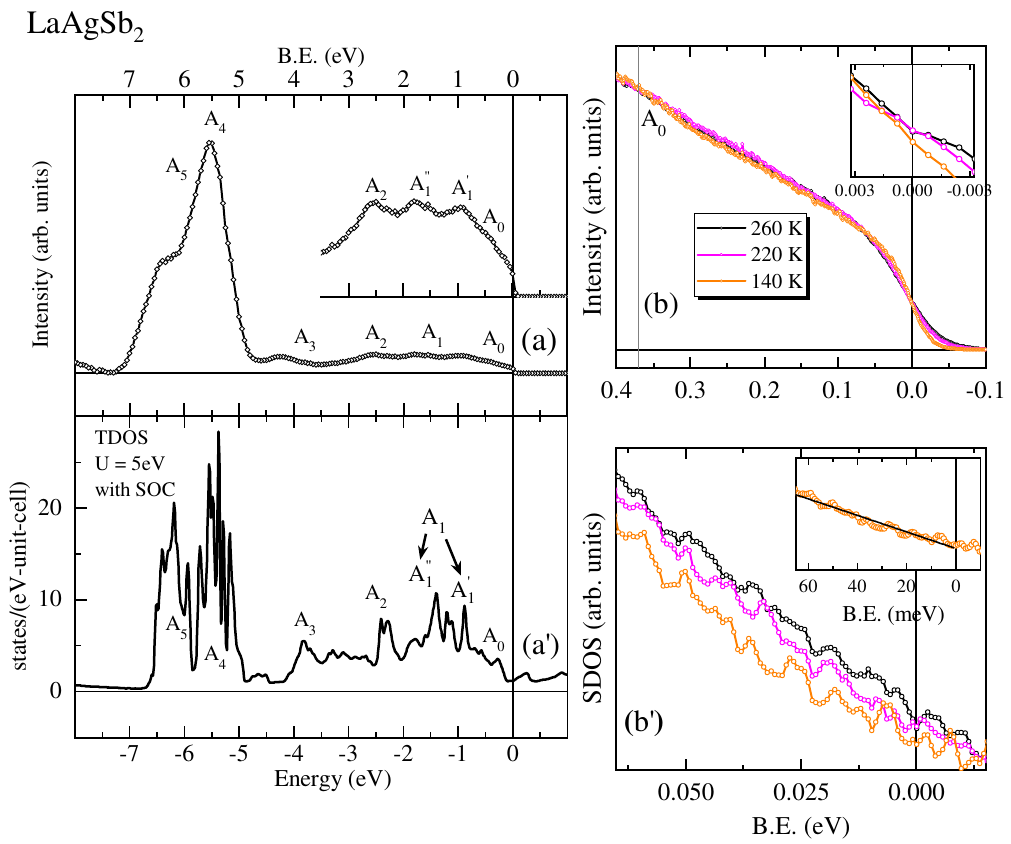}
\caption{ For LaAgSb$_2$ (a) valence band spectra collected using He-I radiation at 30 K and (a') TDOS with $\textit{U}$ = 5 eV with the inclusion of SOC. Temperature-dependent (b) valence band spectra near $\epsilon_F$ for \textit{T} = Ag; the inset shows the expanded view at the $\epsilon_F$; (b') spectral density of states with high resolution He I incident radiation; the inset shows the linear fitting for SDOS 140 K.}
\label{SDOS}

\end{figure}

Fig. \ref{SDOS}(a) shows the valence band (VB) spectrum collected using He-I radiation. The six spectral features ($A_0$ to $A_5$) previously observed in the VB spectum collected using Al K$\alpha$ radiation (Fig. \ref{Valence}(a)) are clearly observed here as well. However, as a result of comparatively higher energy resolution of He-I radiation compared to Al K$\alpha$, the broad $A_1$ feature resolves into two distinct components, $A'_1$ and $A''_1$, as highlighted in the inset of Fig. \ref{SDOS}(a). A comparison of the experimental He-I spectrum with the calculated density of states (Fig. \ref{SDOS}(a')) using $\textit{U}$ = 5 eV including the SOC demonstrates that these two finely resolved features align well with our $\text{DFT}+U$ results. This observation further confirms the energy position of the Ag 4\textit{d} states within the VB of $\text{LaAgSb}_2$.

To understand the behaviour of states close to $\epsilon_F$, let us now focus on the temperature-dependent evolution of states near $\epsilon_F$, investigated using high-resolution He-I spectra, shown in Fig.\ref{SDOS}(b). In case of Al K$_\alpha$ radiation, Fig. \ref{Valence}(a), there is no significant change near the Fermi level (feature A$_0$, ~ 0.37 eV) with temperature for LaAgSb$_2$. Hence all the He-I spectra were normalised around 0.37 eV below $\epsilon_F$. We observe finite DOS at $\epsilon_F$ all the temperatures. To understand the evolution of states close to $\epsilon_F$, spectral density of states (SDOS) was plotted, Fig. \ref{SDOS}(b'). The SDOS was obtained by dividing the experimental spectra by the Fermi-Dirac function for the respective temperatures. It is interesting to note that for T $>$ $T_{C1}$, close to $\epsilon_F$, the SDOS remains almost the same, while for T$<$ $T_{C2-com}$, we observe a slight decrement in the SDOS. The SDOS shows linear energy dependence, close to $\epsilon_F$ as shown in the inset of Fig. \ref{SDOS}(b').

%It is interesting to note that for T $>$ $T_{C1}$, close to $\epsilon_F$, the SDOS remains almost the same, while for T$<$ $T_{C2-com}$, we observe a slight decrement in the SDOS $\epsilon_F$.

In the literature, few angle‑resolved photoemission spectroscopy (ARPES) studies\cite{ARPES_shi2016, ARPES_rosmus_LAS, ARPES_rosmus_LCS} have been done on \textit{T} = Ag compound, but with no clear picture on the CDW mechanism. Shi et al.\cite{ARPES_shi2016} observed a Dirac cone like band structure with nested Fermi surface segments and linked these to the CDW ordering. However, subsequent high‑resolution ARPES at 18 K, well below both CDW transitions, did not resolve a sharp CDW gap in any band within a $\sim$10 meV energy resolution\citep{ARPES_rosmus_LCS}. They found an excellent agreement between ARPES spectra and density functional theory (DFT) calculations without inclusion of CDW structure, over a wide energy range, suggesting that the CDW‑induced reconstruction of the electronic structure is weak.

The decrement in SDOS close to $\epsilon_F$, Fig. \ref{SDOS}(b'), below the CDW transition temperature is suggestive of the fact that there could be nesting of the Fermi surface due to CDW ordering in the system.

\begin{table}
  
  \caption{Electron counts as obtained in the DFT+\textit{U} calculations using $\textit{U}$ = 5 eV, within the muffin tin sphere for different orbitals. }
  
  \begin{tabular}{c|c|c|c}
  \multicolumn{2}{c|}{LaAgSb$_2$} & \multicolumn{2}{c}{LaCuSb$_2$}\\
  \hline
  & Electron Count &  & Electron Count\\
  \hline
  
  La  5\textit{d}& 0.746& La  5\textit{d}& 0.76\\
  6\textit{s}& 0.082  & 6\textit{s}& 0.083 \\
  6\textit{p}& 0.249  & 6\textit{p}& 0.26 \\
  4\textit{f}& 0.091 & 4\textit{f}& 0.09\\
  Ag 4\textit{d}& 9.237 & Cu 3\textit{d}& 9.32 \\
  5\textit{s}& 0.419 & 4\textit{s}& 0.43 \\
  5\textit{p}& 0.379 & 4\textit{p}& 0.39\\
  4\textit{f}& 0.0184 & 4\textit{f}& 0.01\\
  Sb1 5\textit{s}& 1.36& Sb1 5\textit{s}& 1.33\\
  5\textit{p}& 1.59 & 5\textit{p}& 1.51\\
  5\textit{d}& 0.063 & 5\textit{d}& 0.059\\
  4\textit{f}& 0.011 & 4\textit{f}& 0.01\\
  Sb2 5\textit{s}& 1.307&Sb2 5\textit{s}& 1.26\\
  5\textit{p}& 1.79 &  5\textit{p}& 1.72 \\
  5\textit{d}& 0.066 &  5\textit{d}& 0.073\\
  4\textit{f}& 0.020 & 4\textit{f}& 0.021\\
       
  \hline
  
  \end{tabular}
  \label{II}
   
\end{table}

\subsection{Core Level Spectra}

\begin{figure}

\includegraphics [width = 0.48\textwidth]{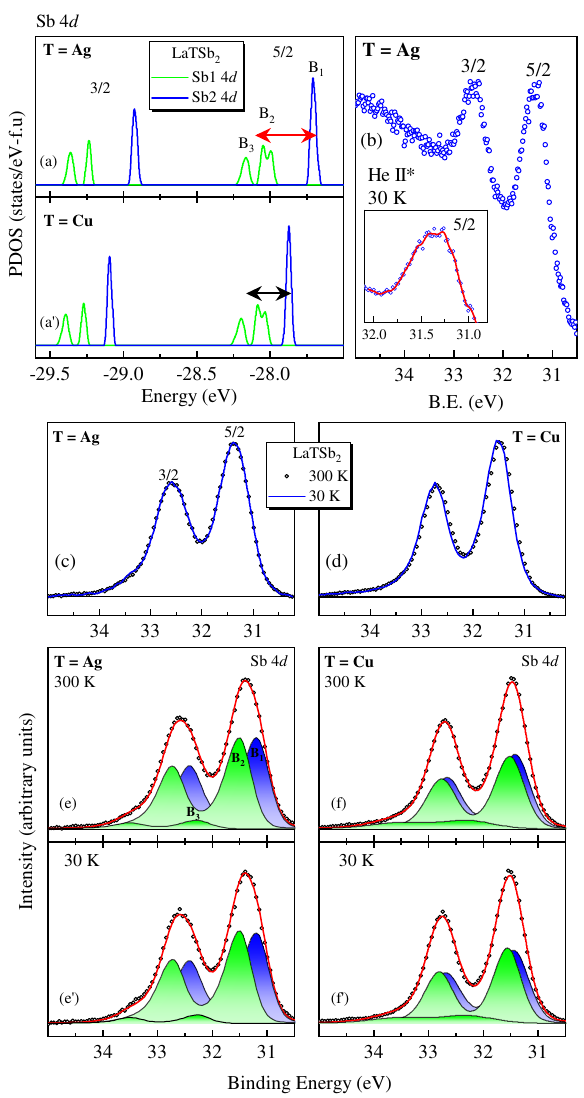}
\caption{Sb 4\textit{d} PDOS for (a) \textit{T} = Ag and (a') \textit{T} = Cu; obtained using $\text{DFT}+U$ calculations, with the inclusion of SOC. The value of $\textit{U}$ is 5 eV. (b) Sb 4\textit{d} spectra for the Ag compound obtained using He II* incident radiation at 30 K; the inset shows the existence of more than one peak corresponding to the spin-orbit split peak of Sb 4\textit{d}$_{5/2}$, the red line is a guide to the eye. Comparison of normalised Sb 4\textit{d} spectra obtained with Al K$_\alpha$ incoming radiation at 300 K and 30 K for (c) \textit{T} = Ag and (d) \textit{T} = Cu. Fitting of the Sb 4\textit{d} spectra for \textit{T} = Ag collected at (e) 300 K and (e') 30 K; for \textit{T} = Cu collected at (f) 300 K and (f') 30 K. In panels (c) and (d) the spectra are normalised by dividing by the area under the curve following the background subtraction.}
\label{Core_Sb4d}

\end{figure}

To obtain information about the electronic structure of the material specific to constituent elements, core level studies were carried out at 300 and 30 K for both the compounds. All the spectra were normalised using the area under the peak after background subtraction.

%\subsection*{Sb 4\textit{d} core level:}

The Sb 4\textit{d} PDOS obtained from $\text{DFT}+U$ calculations for T =Ag and \textit{T} = Cu are shown in Figs. \ref{Core_Sb4d}(a) and (a'), respectively. The energy splitting between spin-orbit coupled peaks is around 1.22 eV. In this figure, corresponding to each spin-orbit peak, we observe three resolved peaks labelled as B$_1$, B$_2$ and B$_3$. The B$_1$ feature arises from the contribution of Sb2, while B$_2$ and B$_3$ originate from Sb1. 

 To understand these behaviours, let us examine the nearest-neighbour bond distances of the Sb atoms, as given in Table \ref{I}. We note that the first nearest neighbour for Sb1 is Sb1 itself, whereas for Sb2, it is a T ion. If we look at the Sb1-Sb1 bond distance, it is also larger than the Sb2-T bond distance, which indicates a stronger hybridisation between Sb2 and T. Thus, the core hole in Sb2 will be efficiently screened by the electron density in T, leading to a feature at lower binding energies. This result is also consistent with valence band results where Sb2 5\textit{p} occupancy is significantly larger than Sb1 5\textit{p} occupancy (refer to Table \ref{II}), i.e., Sb2 valency is more negative than Sb1 valency. This makes it easier for electrons from Sb2 to eject comparatively easier during the photoemission process, giving rise to the feature at lower B.E. in comparison to Sb1.

The Sb 4\textit{d} states corresponding to \textit{T} = Ag occurs at higher energy compared to \textit{T} = Cu. The larger radial extent of Ag 4\textit{d} states, relative to Cu 3\textit{d} states, leads to stronger hybridisation between Sb 5\textit{p} and Ag 4\textit{d} states than between Sb 5\textit{p} and Cu 3\textit{d} states. The localised nature of the Cu 3\textit{d} states results in weaker hybridisation and more atom-like behaviour; consequently, the states corresponding to \textit{T} = Cu are stabilised at lower energy compared to those of \textit{T} = Ag. The energy separation between Sb1 and Sb2 states for T= Ag is more as compared to \textit{T} = Cu. This could be because the Sb1 and Sb2 sites are more electronically distinct in \textit{T} = Ag as compared to that of T=Cu compound.

In Fig. \ref{Core_Sb4d}(b), we show Sb 4\textit{d} core level for \textit{T} = Ag collected at 30 K using He II* source. The spin-orbit splitting energy is $\sim$1.2 eV, which is close to the value obtained from the $\text{DFT}+U$ calculations. In this spectrum, we observe a signature of more than one peak corresponding to each of the spin-orbit split peaks of Sb 4\textit{d} states, shown in inset to Fig. \ref{Core_Sb4d}(b), where we show the spectra corresponding to Sb 4\textit{d}$_{5/2}$. Such behaviour is in line with the literature \cite{ARPES_shi2016}. The observation of multiple peaks suggests the evidence of inequivalent Sb sites in the unit cell.

Figs. \ref{Core_Sb4d}(c) and (d) show Sb 4\textit{d} states collected using Al-K$\alpha$ source at 300 K and 30 K. Unlike the spectra collected using He II*, no clear signature of multiple features are evident. This is due to the lower resolution of the spectrum collected using Al K$_{\alpha}$ as compared to those collected using He II* source.

The temperature dependence of 4d core level spectra for the \textit{T} = Ag compound remains almost the same, while for the case of the \textit{T} = Cu compound, a shift towards higher binding energy is observed on reducing the temperature down to 30 K, as shown in Figs. \ref{Core_Sb4d}(c') and (d').

Furthermore, fitting of the core levels was done to gain better understanding of the features. The Sb 4\textit{d} core level spectra were fitted with fixed spin-orbit splitting values of 1.22 eV and 1.24 eV for the cases of \textit{T} = Ag and \textit{T} = Cu, respectively, shown in Figs. \ref{Core_Sb4d} (e),(e') and (f), (f'). Our detailed analysis suggests that better fitting could be achieved only when three peaks, labelled as B$_1$, B$_2$ and B$_3$ were considered for Sb 4\textit{d} spectra. These results are in line with our calculations.

\begin{figure}

\includegraphics [width = 0.48\textwidth]{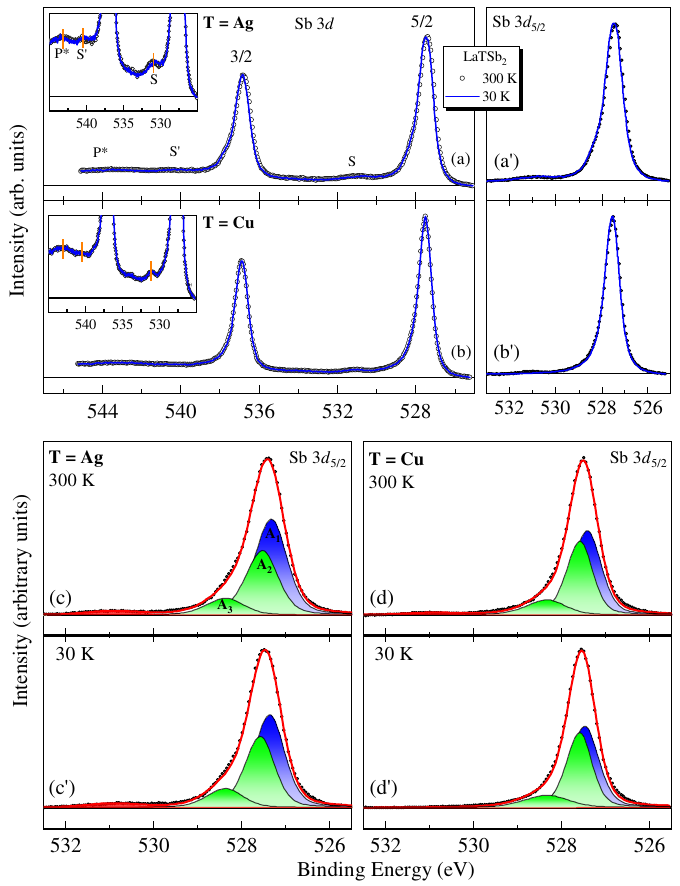}
\caption{Comparison of Sb 3\textit{d} spectra collected under Al K$_\alpha$ incident radiation at 300 K and 30 K for (a) \textit{T} = Ag and (b) \textit{T} = Cu; insets provide enhanced views of distinct features identified other than the principal peak; normalised Sb 3\textit{d}$_{5/2}$ spectra for (a') \textit{T} = Ag and (b') \textit{T} = Cu. Fitting of the Sb 3\textit{d}$_{5/2}$ spectra for \textit{T} = Ag collected at (c) 300 K and (c') 30 K; for \textit{T} = Cu collected at (d) 300 K and (d') 30 K. In panels (a) and (b), the spectra are normalised to the Sb 3\textit{d}$_{5/2}$ peaks. In panels (a') and (b'), the spectra are normalised by dividing by the area under the curve after background subtraction. }
\label{Core_Sb3d}

\end{figure}

In Figs. \ref{Core_Sb3d}(a) and (b), we show temperature-dependent Sb 3\textit{d} core-level spectra for \textit{T} = Ag and Cu, respectively. The spin-orbit splitting energy of Sb 3\textit{d} states is $\sim$ 9.35 eV. Corresponding to each of the spin-orbit split peaks, at the higher binding energy side, we observe a small hump labelled S and S'. One more feature labelled as P* is attributed to the contribution from plasmons\cite{morgan2017metallic}, which is weak in the case of the studied compounds but can be clearly visible in the inset of Figs. \ref{Core_Sb3d}(a) and (b).

It is important to note that, generally around 531 eV, the contribution of surface oxygen occurs. However, in the present case, we can neglect the possibility of surface oxygen because (a) the samples are cleaved \textit{in situ} under ultra-high vacuum conditions and (b) for the surface oxygen peak coming out of O 1s levels, it is expected that no spin-orbit coupled counterpart occurs. However, in the present case, we observe such counterparts, labelled as S and S$'$, Figs. \ref{Core_Sb3d}(a) and (b).

It is noteworthy that Sawani et al.\cite{CeCuSb2_Maiti} investigated the core-level electronic structure of $\text{CeCuSb}_2$ using hard X-ray photoemission spectroscopy (HAXPES). Specifically, their $\text{Sb } 3d_{5/2}$ spectra revealed four features in the binding energy range of 526–533 eV. Based on energy-dependent intensity variations, they classified these into surface and bulk components, with two peaks allocated to each of the inequivalent Sb sites (Sb1 and Sb2). While a definitive assignment is difficult from the current data, the features S and S$'$ likely originate from surface states. Conclusively resolving the physical origin of these features will require systematic energy-dependent HAXPES measurements. Additionally, with a decrease in temperature, a small shift towards high binding energy is observed for the case of the \textit{T} = Ag system, while for the case of the \textit{T} = Cu system, the peak position remains almost the same, as shown in Figs. \ref{Core_Sb3d} (a') and (b').

In Figs. \ref{Core_Sb3d}(c, c', d and d'), we show the fitting of Sb 3\textit{d}$_{5/2}$ core levels for both the samples collected at 300 and 30 K. Our results show that a satisfactory fit can not be achieved when the Sb 3\textit{d}$_{5/2}$ core level spectra was fitted using two peaks, the FWHM for higher binding energy side peak was almost double in comparison to lower binding energy peak, that is unusual. To resolve this issue, the spectra were fitted using three peaks, starting with all the peaks having similar FWHM, labelled as A$_1$, A$_2$ and A$_3$, Fig. \ref{Core_Sb3d}(c). This observation is consistent with the fitting of Sb 4\textit{d} spectra. Separation between A$_1$ and A$_2$ peaks for the case of \textit{T} = Ag is low in comparison to \textit{T} = Cu, that is consistent with $\text{DFT}+U$ results and Sb 4\textit{d} fittings; refer to Tables \ref{III} and \ref{IV} in supplementary material.

Additionally, our results show that the combined area under the peaks of A$_2$ and A$_3$ arising from Sb1 is almost equal to the area under the peak of A$_1$ due to Sb2, where A$_1$ $ :$ (A$_2$ $+$ A$_3$) comes out to be $\sim$ 1.0(1) and 0.95(1) for the case of \textit{T} = Ag and \textit{T} = Cu compounds, Table \ref{IV} in supplementary data.

It is important to note here that in a recent report based on hard x-ray photoemission studies \cite{chainani2025_LAS}, in the Sb 3\textit{d} core level spectra, two features were clearly observed around the positions of peaks A$_1$ and A$_3$. The 220 K core level spectrum aligns well with our 300 K data, i.e., above the CDW transition temperatures. With the decrease in temperature, i.e., below CDW transition temperatures, they found an increase in the intensity of the feature around A$_3$.

In the present study, with a decrease in temperature, we observe that the area under the curve for the A$_3$ feature slightly increases in the case of the \textit{T} = Ag sample, whereas it decreases slightly for \textit{T} = Cu. Our findings suggest that the A$_3$ feature that is related to the Sb1 square net, plays a significant role in CDW formation in the \textit{T} = Ag compound.

\begin{figure}

\includegraphics [width = 0.48\textwidth]{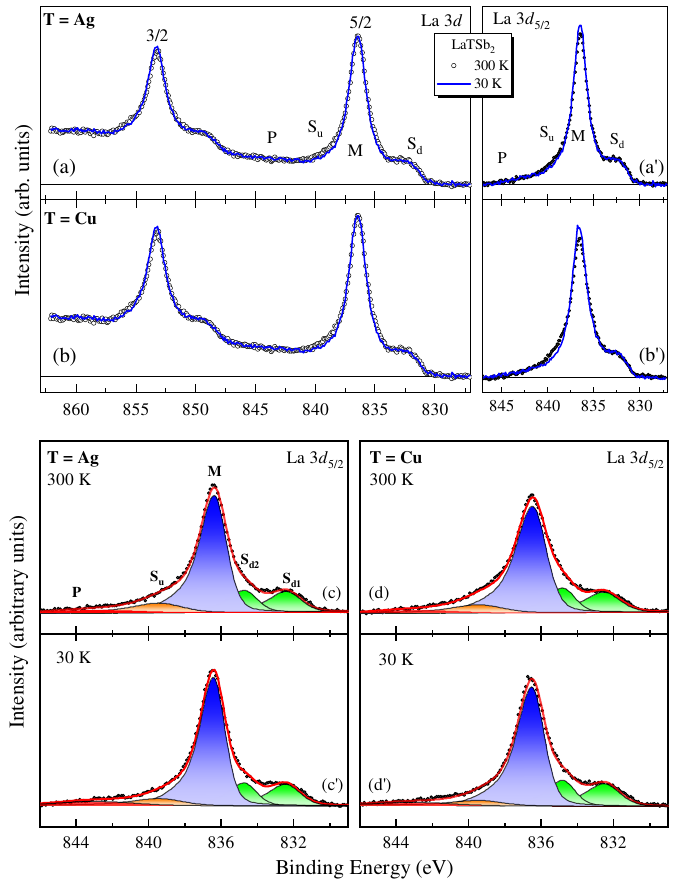}
\caption{Comparison of La 3\textit{d} spectra obtained with Al K$_\alpha$ incident radiation at 300 K and 30 K for (a) \textit{T} = Ag and (b) \textit{T} = Cu; normalised La 3\textit{d}$_{5/2}$ spectra for (a') \textit{T} = Ag and (b') \textit{T} = Cu. Fitting of the La 3\textit{d}$_{5/2}$ spectra for \textit{T} = Ag collected at (c) 300 K and (c') 30 K; for \textit{T} = Cu collected at (d) 300 K and (d') 30 K. In panels (a) and (b), the spectra are normalised to the La 3\textit{d}$_{5/2}$ peaks. In panels (a') and (b'), the spectra are normalised by dividing by the area under the curve after background subtraction. }
\label{Core_La3d}

\end{figure}

In Figs. \ref{Core_La3d}(a) and (b), we show a comparison of the La 3\textit{d} core level spectra collected at 300 K and 30 K for \textit{T} = Ag and Cu compounds. In the spectra collected for both the compounds, Figs. \ref{Core_La3d}(a') and (b'), we observe two distinct features labelled as S$_d$ and M and a tail at the higher binding energy side with two features labelled as S$_u$ and P. All these four features are attributed to well screened/shake-down satellite, a poorly screened main peak, a shake-up satellite and a plasmon peak, respectively. Such features are also observed in La metal\cite{schlapbach1981xps} and LaAl$_2$\cite{slebarski2022_LaAl2} compounds.

With a decrease in temperature, we observe an increment in the intensity of the feature M for both the samples. To understand the behaviours of these features, core level fittings were carried out for both the temperatures, Figs. \ref{Core_La3d}(c-d') in the binding energy range 827.24 - 847.14 eV eV. In this range, it was found that four peaks were not sufficient to fit the data; instead, five peaks were required. For better fitting of feature S$_d$, in our case, two peaks, namely S$_{d1}$ and S$_{d2}$, were required.

The two features, S$_{d1}$ and S$_{d2}$, corresponding to the shake-down satellite, have been attributed to the charge transfer from the inequivalent Sb atoms. The area under the curve corresponding to these features was constrained to a 1:1 ratio based on the occupancies of these atoms in the unit cell. It is important to note that in the La 3\textit{d}$_{5/2}$ core levels of the LaTe$_3$ compound\citep{sarkar2020_LaTe3}, around 832 eV, two features were used to fit the core level spectrum. In the case of LaAl$_2$ system, three features corresponding to shake down, poorly screened and shake up features have been indentified \citep{slebarski2022_LaAl2}.

During the photoemission process, in the La 3\textit{d} core level, after a core hole is created, three final states are possible, as observed in the case of LaAl$_2$\citep{slebarski2022_LaAl2}. The feature M originates due to the direct photoemission process of the 3\textit{d} electron, leading to the final state $3d^9 4f^0$. The feature S$_d$ arises when there is screening of the core hole due to the transfer of an electron from the states close to the $\epsilon_F$ to the 4f states, leading to a $3d^9 4f^1$ final state. Such a transfer becomes possible when the 4f states lower their energy below the $\epsilon_F$ due to the core hole created.

The feature S$_u$ arises due to another final state in which an electron from the 5d6s bands gets excited into an unoccupied 4f state (above the $\epsilon_F$) during the photoemission process\cite{slebarski2022_LaAl2}. Because of such occupation of the excited electron, the 4f states lower their energy below the $\epsilon_F$. For better visualisation, we have compared the La 3\textit{d} spectra of La metal, LaAl$_2$, and LaAgSb$_2$ in the supplementary data [Fig. S1].

The results of the fitting are shown in Table \ref{V} of the supplementary data. In \textit{T} = Ag, with a decrease in temperature, the probability of the processes associated with feature (a) S$_d$ remains almost the same, (b) M and P exhibits a small increment, and (c) S$_u$ shows a decrement. At low temperatures, in the CDW phase, an increment in the area under the peak M could be due to reorganisation of the conduction electrons due to partial formation of the gap at the Fermi surface thereby leading to the decrement in the effective screening of the core hole. The decrement in the area under the peak S$_u$ could be due to the decrement in the density of states at the $\epsilon_F$ thereby leading to the reduction in the low energy excitations.

For \textit{T} = Cu, at low temperatures, the relative probability of the processes associated with feature (a) S$_d$ and M exhibit a small increment, (b) S$_u$ exhibits a decrement, and (c) P remains almost the same. At low temperature, the increment in area under the peak M and S$_d$ could be related to the small increase observed in DOS at the $\epsilon_F$, shown in Fig. \ref{Valence} (b). As the DOS at the $\epsilon_F$ increases, the probability of the process related to screening increases, which makes the probability of process related to the shake-up satellite (S$_u$) to decrease.

\begin{figure}

\includegraphics [width = 0.48\textwidth]{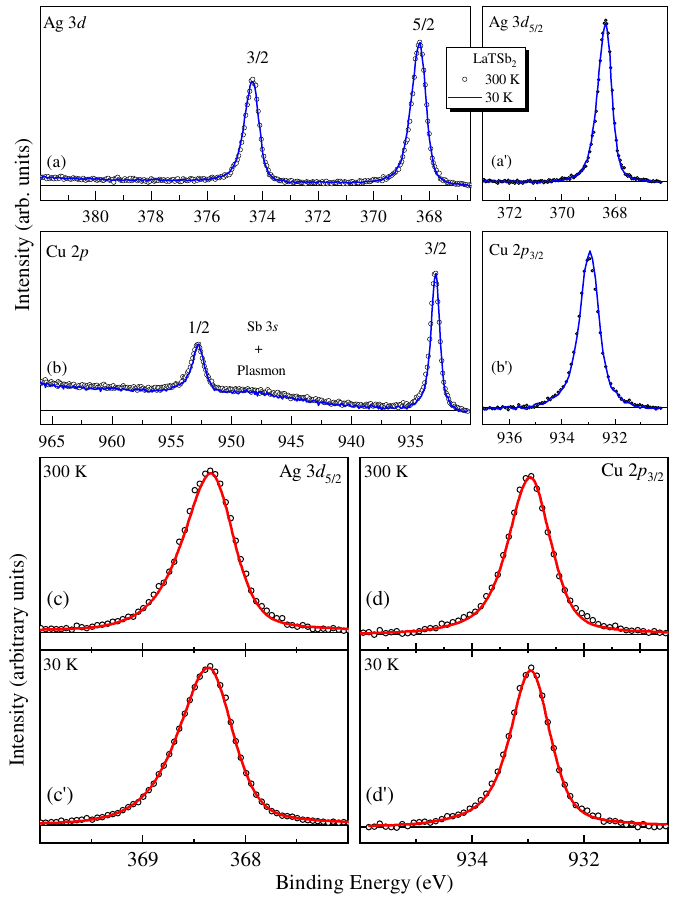}
\caption{Comparison of spectra obtained with Al K$_\alpha$ incident radiation at 300 K and 30 K for (a) Ag 3\textit{d}, \textit{T} = Ag and (b) Cu 2\textit{p}, \textit{T} = Cu; normalised spectra for (a') Ag 3\textit{d}$_{5/2}$, \textit{T} = Ag and (b') Cu 2\textit{p}$_{3/2}$, \textit{T} = Cu. Fitting of Ag 3\textit{d} $_{5/2}$ spectra for \textit{T} = Ag collected at (c) 300 K and (c') 30 K; fitting of Ag 3\textit{d} $_{5/2}$ spectra for Cu at (d) 300 K and (d') 30 K. In the spectra shown in panels (a) and (b), the normalisation is based on the Ag 3\textit{d}$_{5/2}$ and Cu 2\textit{p}$_{3/2}$ peaks, respectively, while the spectra shown in panels (a') and (b') are normalised by dividing by the area under the curve after background subtraction.}
\label{Core_Ag3d_Cu2p}

\end{figure}
%The Ag 3\textit{d} and Cu 2\textit{p} spectra for Ag and Cu metals displayed in (a) and (b) have been obtained from Kulbacki et al.\cite{kulbacki2024analysis} and Dean et al.\cite{dean2023analysis}, respectively.

Temperature-dependent Ag 3\textit{d} spectra and normalised Ag 3\textit{d}$_{5/2}$ core-level spectra are shown in Figs. \ref{Core_Ag3d_Cu2p}(a) and (a'), respectively. The core level spectra was fitted using a single peak for both the temperatures, Figs. \ref{Core_Ag3d_Cu2p} (c) and (c'). Additionally, unlike pure Ag metal\cite{kulbacki2024analysis}, the contribution of plasmon peaks that is expected to be observed around 372 eV and 378 eV is almost negligible in the \textit{T} = Ag compound. It is hard to comment on the valence state of Ag in the system only based on Ag 3\textit{d} spectra, as the Ag 3\textit{d} peak position varies minimally ($\sim$ 0.5 eV) with the change in oxidation state\cite{hoflund2000surface}. For that, more detailed studies of the 3\textit{d} peak along with Auger peaks will be necessary.

In Figs. \ref{Core_Ag3d_Cu2p}(b) and (b'), we show a comparison of the Cu 2\textit{p} and normalised Cu 2\textit{p}$_{3/2}$ core levels collected at 300K and 30K, respectively. The Cu 2\textit{p} spectrum can be used as a fingerprint to determine the nominal Cu valency in the system. As per the available literature, a clean narrow peak without satellites indicates Cu$^{+1}$, whereas Cu$^{+2}$ shows a broad main peak with a satellite structure $\sim$ 9 eV towards the higher B.E. side\cite{ghijsen1988electronic, chainani2014electronic}. Our narrow Cu 2\textit{p} spectra without any satellite features suggest the monovalency of Cu in LaCuSb$_2$\citep{CeCuSb2_Maiti, chainani2014electronic}. The fitting of the core level was carried out using a single peak for both temperatures, Figs. \ref{Core_Ag3d_Cu2p} (d) and (d').

\begin{figure}

\includegraphics [width = 0.50\textwidth]{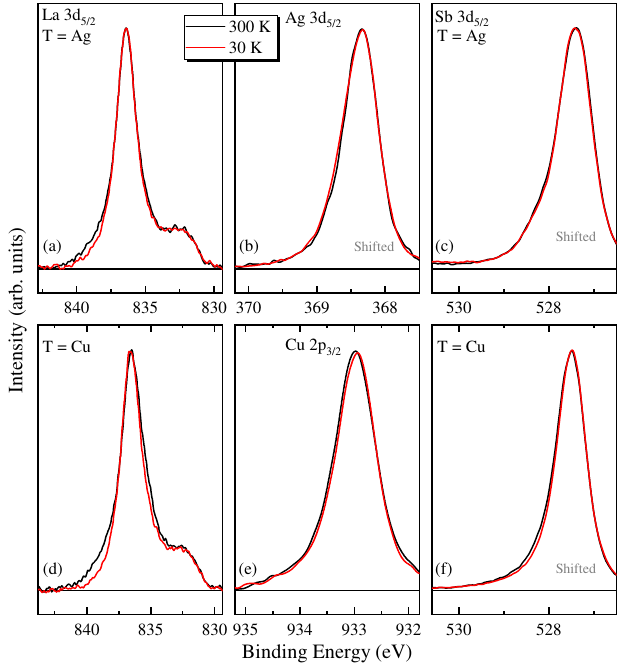}
\caption{Comparison of normalised temperature-dependent core level (a) La 3\textit{d}$_{5/2}$, (b) Ag 3\textit{d}$_{5/2}$, and (c) Sb 3\textit{d}$_{5/2}$ for \textit{T} = Ag;(a) La 3\textit{d}$_{5/2}$, (b) Cu 2\textit{p}$_{3/2}$, and (c) Sb 3\textit{d}$_{5/2}$; for \textit{T} = Cu. Here, normalisation is done at the peak maximum following the background subtraction. After normalisation Ag 3\textit{d}$_{5/2}$ spectra and Sb 3\textit{d}$_{5/2}$ spectra have been shifted to match the lower B.E. side.}
\label{Comp_temp}

\end{figure}

To visualise the temperature-dependent variation, all the spectra were normalised to match the peak maximum. The low-temperature spectra of Ag and Sb 3\textit{d}$_{5/2}$ were shifted to match the lower binding energy side of the 300 K spectra, as shown in Figs. \ref{Comp_temp} (a-f). In the case of \textit{T} = Cu, at 30 K, the width of the most intense peak is found to decrease as compared to the 300 K spectra due to thermal effects, while in the case of the spectrum of the \textit{T} = Ag compound, the peak width remains almost the same. In fact, for the case of Ag 3\textit{d}$_{5/2}$ the fitting results show a slight increase in the FWHM as we go from 300 K to 30 K, as shown in Table \ref{VI}. Such counter-intuitive behaviour in the LaAgSb$_2$ could be due to the stabilisation of the CDW phase, where the emergence of inequivalent electronic environments at low temperatures introduces additional broadening that compensates for the thermal narrowing.

\section{Summary}

Sb square net-based Dirac semimetallic ternary antimonides LaTSb$_2$ (\textit{T} = Cu, Ag) are investigated using magnetotransport, photoemission, and density functional theory (DFT) based ab initio calculations. Both the compound exhibits positive TCR, direction-dependent linear MR and deviation from Fermi liquid behaviour. Whereas, only \textit{T} = Ag compounds exhibit effects of charge density waves (CDW) in resistivity data. Further extensive temperature-dependent MR measurements for \textit{T} = Ag system show a slope change from linear to quadratic as the CDW critical temperature approaches. Kohler's scaling suggests that single-band semiclassical transport is insufficient to explain the transport behaviour in \textit{T} = Ag system. DFT + U calculations are required to reproduce the experimental valence band electronic structure. The core-level spectra reveal the presence of inequivalent Sb sites in the Sb 3\textit{d}, 4\textit{d} and La 3\textit{d} spectra. Sb 4\textit{d} core-level features were reproduced with full potential $\text{DFT}+U$ calculations. The $T = \text{Cu}$ system shows predicted temperature-dependent core-level sharpening due to reduced phonon scattering, the $T = \text{Ag}$ compound shows an unusual low-temperature broadening in the CDW phase. For the case of \textit{T} = Ag system, Sb 3\textit{d} core level analysis reveals the involvement of Sb square net in the CDW phenomena. Comprehensive UPS studies (He-I spectra) on the LaAgSb$_2$ sample reveal a slight decrease in spectral DOS near the Fermi level ($\epsilon_F$) at low temperatures in the CDW phase. We believe that the present core-level investigation will provide a valuable framework for understanding the electronic structures of CDW and non-CDW systems, elucidating the connection between local electronic environments and CDW formation, and establishing reference core-level spectral features that can aid future spectroscopic investigations of related Sb-based materials.

\section{Acknowledgement}

One of the authors (A.T.) acknowledges the support of the Department of Atomic Energy, Government of India, under Project Identification No. RTI 4015.

\bibliography{Manuscript}

\onecolumngrid

\section*{Supplementary Material}

\begin{figure*}[!ht]
\centering
\includegraphics[width=0.75\textwidth]{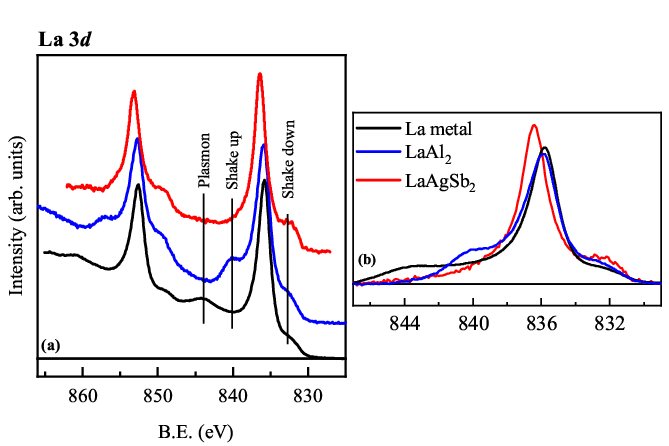}
\caption{Comparison of (a) La 3\textit{d} core level spectra and (b) normalised La 3\textit{d}$_{5/2}$ spectra for La metal (taken from Morgan et al.\cite{morgan2024core}), LaAl$_2$ (data received after request, courtesy of Prof. Andrzej Ślebarski.\cite{slebarski2022_LaAl2}), and LaAgSb$_2$.}
\label{Sup}
\end{figure*}

\begin{table*}[!ht]
\centering
\setlength{\tabcolsep}{2.0pt}
\renewcommand{\arraystretch}{1.5}
\caption{Results from Sb 4\textit{d} core-level fitting}
\begin{tabular}{l|l|c|c|c|c|c|c}
\hline
&& \multicolumn{1}{c}{} & \multicolumn{1}{c}{300 K} &  & \multicolumn{1}{c}{} &\multicolumn{1}{c}{30 K}&\\
&{Peak} & {Position (eV)} & {Area} & {FWHM (eV)} & {Position (eV)} & {Area} & {FWHM (eV)} \\
\hline
    LaAgSb$_2$& B1 & 31.192 & 0.565 & 0.530 & 31.196 & 0.561 & 0.530 \\
     &B2 & 31.507 & 0.553 & 0.520 & 31.504 & 0.563 & 0.520 \\
     &B3 & 32.286 & 0.059 & 0.583 & 32.270 & 0.051 & 0.510 \\
     \hline
    LaCuSb$_2$&B1 & 31.417 & 0.452 & 0.616 & 31.448 & 0.463 & 0.643 \\
     &B2 & 31.517 & 0.438 & 0.616 & 31.560 & 0.429 & 0.578 \\
     &B3 & 32.309 & 0.095 & 1.131 & 32.310 & 0.088 & 1.130 \\
\hline
\end{tabular}
\label{III}
\end{table*}

\begin{table*}[!ht]
\centering
\setlength{\tabcolsep}{2.0pt}
\renewcommand{\arraystretch}{1.5}
\caption{Results from Sb 3\textit{d}$_{5/2}$ core-level fitting.}
\begin{tabular}{l|l|c|c|c|c|c|c}
\hline
&& \multicolumn{1}{c}{} & \multicolumn{1}{c}{300 K} &  & \multicolumn{1}{c}{} &\multicolumn{1}{c}{30 K}&\\
&{Peak} & {Position (eV)} & {Area} & {FWHM (eV)} & {Position (eV)} & {Area} & {FWHM (eV)} \\
\hline
    LaAgSb$_2$& A1 & 527.321 & 0.552 & 0.831 & 527.356 & 0.509 & 0.791 \\
     &A2 & 527.524 & 0.373 & 0.833 & 527.573 & 0.376 & 0.761 \\
     &A3 & 528.36  & 0.114 & 0.968 & 528.362 & 0.123 & 0.929 \\
     &S  & 530.94  & 0.03  & 1.38  & 530.94  & 0.03  & 1.38  \\
     \hline
    LaCuSb$_2$&A1 & 527.408 & 0.427 & 0.733 & 527.468 & 0.417 & 0.741 \\
     &A2 & 527.577 & 0.343 & 0.67  & 527.585 & 0.341 & 0.654 \\
     &A3 & 528.329 & 0.105 & 0.999 & 528.324 & 0.1   & 1.157 \\
     &S  & 530.94  & 0.014 & 1.38  & 530.94  & 0.001 & 1.38  \\
\hline
\end{tabular}
\label{IV}
\end{table*}

\begin{table*}[!ht]
\centering
\setlength{\tabcolsep}{2.0pt}
\renewcommand{\arraystretch}{1.5}
\caption{Results from La 3\textit{d}$_{5/2}$ core-level fitting.}
\begin{tabular}{l|l|c|c|c|c|c|c}
\hline
&& \multicolumn{1}{c}{} & \multicolumn{1}{c}{300 K} &  & \multicolumn{1}{c}{} &\multicolumn{1}{c}{30 K}&\\
&{Peak} & {Position (eV)} & {Area} & {FWHM (eV)} & {Position (eV)} & {Area} & {FWHM (eV)} \\
\hline
LaAgSb$_2$&S2 & 832.417 & 0.149 & 1.928 & 832.453 & 0.151 & 1.874 \\
     &S1 & 834.736 & 0.149 & 1.697 & 834.747 & 0.151 & 1.649 \\
     &M  & 836.391 & 0.742 & 1.593 & 836.429 & 0.758 & 1.485 \\
     &L  & 839.447 & 0.119 & 3.034 & 839.450 & 0.085 & 3.030 \\
     &P  & 843.000 & 0.040 & 5.630 & 843.000 & 0.052 & 5.022 \\
     \hline
LaCuSb$_2$&S2 & 832.555 & 0.160 & 2.123 & 832.526 & 0.168 & 2.010 \\
     &S1 & 834.821 & 0.160 & 1.630 & 834.829 & 0.168 & 1.637 \\
     &M  & 836.487 & 0.757 & 1.783 & 836.528 & 0.766 & 1.619 \\
     &L  & 839.447 & 0.100 & 3.034 & 839.450 & 0.065 & 3.030 \\
     &P  & 843.000 & 0.035 & 5.022 & 843.000 & 0.039 & 5.022 \\
\hline
\end{tabular}
\label{V}
\end{table*}

\begin{table*}[!ht]
\centering
\setlength{\tabcolsep}{2.0pt}
\renewcommand{\arraystretch}{1.5}
\caption{Results from Ag 3\textit{d}$_{5/2}$ and Cu 2\textit{p}$_{3/2}$ core-level fitting.}
\begin{tabular}{l|l|c|c|c|c|c|c}
\hline
&& \multicolumn{1}{c}{} & \multicolumn{1}{c}{300 K} &  & \multicolumn{1}{c}{} &\multicolumn{1}{c}{30 K}&\\
&{Peak} & {Position (eV)} & {Area} & {FWHM (eV)} & {Position (eV)} & {Area} & {FWHM (eV)} \\
\hline
LaAgSb$_2$ & Ag 3\textit{d}$_{5/2}$ & 368.33 & 0.51 & 1.07 & 368.36 & 0.55 & 1.15 \\
     \hline
LaCuSb$_2$& Cu 2\textit{p}$_{3/2}$ & 932.96 & 0.85 & 1.20 & 932.95 & 0.8 & 1.17 \\
\hline
\end{tabular}
\label{VI}
\end{table*}

\end{document}